# Narwhal-shaped Wavefunctions Enabling Three-dimensional Sub-diffraction-limited Dielectric Photonics

Wen-Zhi Mao[1,*], Hong-Yi Luan[1,*], Ren-Min Ma[1,2,3,†]

[1]State Key Laboratory for Mesoscopic Physics and Frontiers Science Center for Nano-optoelectronics, School of Physics, Peking University, Beijing, China
[2]Peking University Yangtze Delta Institute of Optoelectronics, Nantong, Jiangsu, China
[3]National Biomedical Imaging Center, Peking University, Beijing, China

Field localization, characterized by mode volume, is central to optics, photonics, and all light–matter interactions. Smaller mode volumes amplify the electric field per photon, enhancing spontaneous emission, strengthening nonlinear optical effects, and enabling strong coupling in cavity quantum electrodynamics. However, in lossless dielectric systems, the diffraction limit has long been considered an unbreakable barrier to light confinement. Here, we uncover a novel class of wavefunctions—narwhal-shaped wavefunctions—and reveal their pivotal role in enabling extreme light confinement in lossless dielectrics across all spatial dimensions. Through rigorous theoretical analysis, simulations, and experimental validation, we propose and realize a three-dimensional singular cavity supported by these wavefunctions, achieving an ultrasmall mode volume of $5\times10^{-7}\ \lambda^3$ ($\lambda$: free-space wavelength). Our findings open new frontiers for unprecedented control over light–matter interactions at the smallest possible scales.

*Introduction*—In 1927, Dirac's revolutionary quantization of the electromagnetic field redefined our understanding of light, demonstrating that each electromagnetic mode can be treated as a quantum harmonic oscillator confined within a finite cavity [1]. By defining a quantization volume, electromagnetic modes become discrete and normalized, with their energy quanta directly interpreted as photons. This framework not only bridges the gap between quantized field modes and photonic excitations but also establishes the critical role of mode volume: smaller mode volumes amplify the electric field per photon, thereby enhancing light–matter interactions. These principles form the cornerstone of cavity quantum electrodynamics [2–6], drive the evolution of advanced photonic technologies [7–40], and fuel the progression of modern quantum optics [41–43].

Despite these advances, photonic devices continue to lag behind their electronic counterparts in terms of integration density and spatial resolution—a disparity rooted in the optical diffraction limit. In the visible and near-infrared regimes, the wavelength of photons is approximately three orders of magnitude larger than the de Broglie wavelength of electrons in electronic devices. This discrepancy imposes a fundamental constraint: the smallest achievable photonic mode volume is roughly nine orders of magnitude larger than the corresponding volume for electrons—nearly a billion times greater. Plasmonics has provided a means to overcome this diffraction limit, enabling significant breakthroughs in sensing, imaging, and on-chip photonics [7–19]. However, the unavoidable ohmic losses of metals remain a severe bottleneck, limiting their performance and scalability [17, 44–45].

Achieving extreme photon confinement in a lossless system is essential for advancing photonic integration and imaging capabilities. Such a breakthrough would enable transformative applications that demand precise nanoscale control of light, including the optical observation of biomolecular structures such as DNA and the development of large-scale photonic integrated circuits with significantly enhanced processing speeds and efficiencies. Overcoming this challenge requires addressing a key limitation: achieving sub-diffraction-limited confinement in a lossless dielectric system.

Recent theoretical [20], simulation, and experimental progress [20–25] has introduced an emerging frontier in photonics: sub-diffraction-limited dielectric photonics. The singular dispersion equation, discovered for lossless dielectric materials, reveals power-law divergent wavefunctions capable of achieving sub-diffraction confinement [20]. However, these wavefunctions remain experimentally unverified, and their confinement is fundamentally restricted to two spatial dimensions, falling short of achieving full three-dimensional confinement. Resolving this challenge presents a compelling opportunity to advance photonics, with the potential to unlock unprecedented capabilities in light manipulation and device performance.

In this work, we propose and experimentally demonstrate a three-dimensional singular dispersion equation that enables sub-diffraction-limited electromagnetic field confinement within dielectric singular cavities. Our cavity design integrates a dielectric biconical antenna within a twisted lattice architecture, producing distinctive narwhal-shaped wavefunctions. These wavefunctions exhibit a power-law divergence at their core, combined with a rapid exponential decay, allowing the field to be sharply confined in all three spatial dimensions and significantly reducing the mode volume (Fig. 1). Through comprehensive full-wave electromagnetic simulations and direct experimental validation, we achieve an unprecedented mode volume of just $5\times10^{-7}$ $\lambda^3$ (where $\lambda$ is the free-space wavelength), surpassing the previously reported two dimensional benchmark by nearly three orders of magnitude [20, 23-25].

---

* These authors contributed equally to this work.
†Contact author: renminma@pku.edu.cn

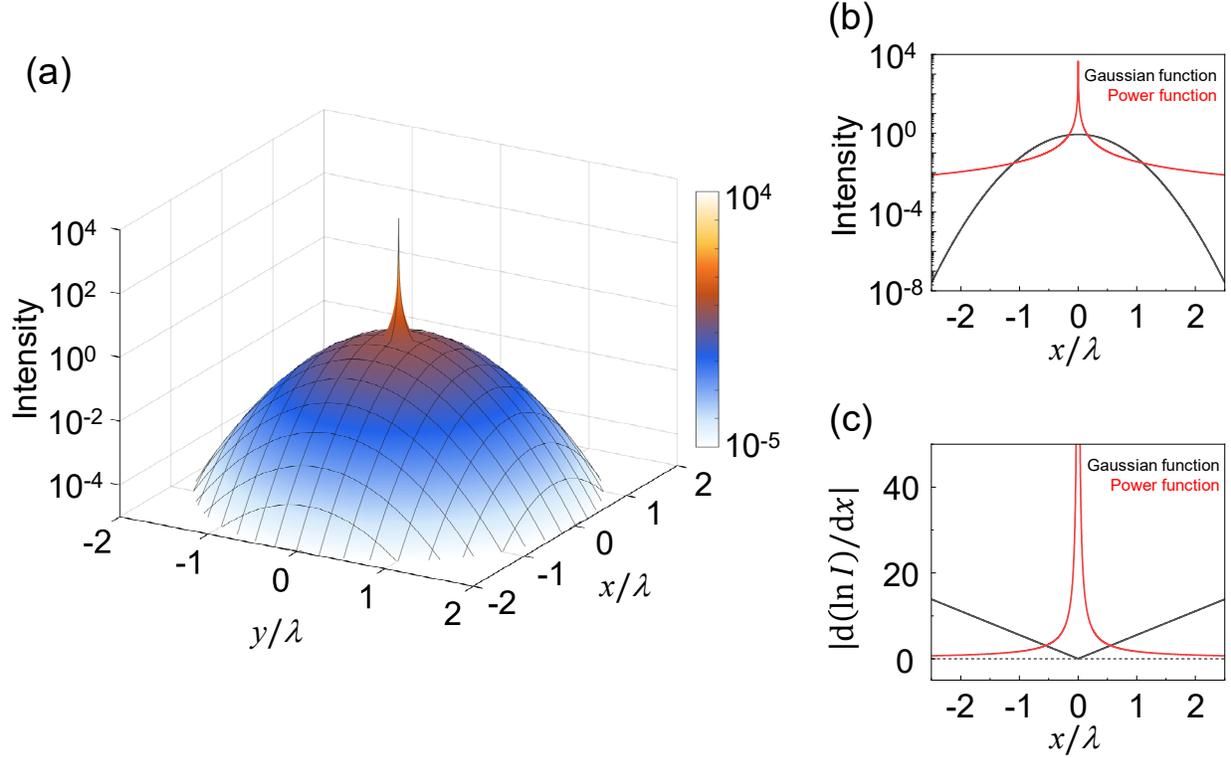

FIG. 1. Narwhal-shaped wavefunction. The mode volume describes how far an optical mode extends in space; minimizing it requires rapid decay of the electric field. We propose a hybrid wavefunction that transitions from a power-law profile near its peak intensity to an exponential tail. Although a Gaussian function decays rapidly at large distances, its relatively gentle slope near the peak impedes field attenuation. By contrast, a power-law profile steepens as one moves closer to the center, producing a faster drop-off near the maximum. Merging these behaviors ensures swift decay throughout the spatial domain, significantly reducing the mode volume. Because its shape resembles a narwhal's head, we term it the narwhal-shaped wavefunction. (a) Log-scale intensity distribution of a narwhal-shaped wavefunction. (b) Log-scale comparison between a Gaussian function and a power function. Because the power-law function diverges when integrated over the entire spatial domain, it cannot serve as an eigen-wavefunction on its own. Here, it is normalized within a circular region of diameter $2.5\lambda$, where $\lambda$ is the free-space wavelength. (c) Relative rate of change of the electric field intensity with respect to position, $\left|\frac{d(\ln I)}{dx}\right|$, which equals $\frac{|x|}{\sigma^2}$ for the Gaussian function and $\frac{2l}{|x|}$ for the power function.

*Three-dimensional singular field*—We find that a dielectric biconical antenna supports a three-dimensional singular field, which rapidly diverges following a power law in all spatial dimensions as it approaches the apex singularity of the conical dielectric structures. Near the singularity (where $k_0 r \ll 1$, with $k_0$ being the free space wavevector and $r$ representing the distance from the apices in spherical coordinates), the eigen-wavefunction is given by $\mathbf{E}_s = C_s r^{-l} \Theta(\theta, \varphi)$, where $C_s$ is a constant, $l$ is a constant between 0 and 1, $\Theta(\theta, \varphi)$ is a function of the spherical coordinate angles $\theta$ and $\varphi$.

This three-dimensional singular field indicates that the electric field varies with the distance $r$ from the singularity following a power law with exponent $l$ in any spatial direction defined by the angles $\theta$ and $\varphi$. This scaling law allows the electric field to vary by several orders of magnitude within a subwavelength range (Fig. 1), serving as a fundamental mechanism for overcoming the optical diffraction limit in dielectric systems.

As $r$ approaches zero, the electric field tends to infinity. This divergence in dielectric systems corresponds to the divergence of the mode wavevector. For any polarization component of the electric field, we can use a position-dependent wavevector by rewriting $E(r,\theta,\varphi) = e^{i\mathbf{k}(r,\theta,\varphi)\cdot\mathbf{r}}$ as $E(r,\theta,\varphi) = e^{i\int \mathbf{k}(r,\theta,\varphi)\cdot d\mathbf{r}}$. This approach allows us to capture the wavevector associated with the power-law behavior and derive the corresponding dispersion equation. The power-function-scaled electric field causes all wavevector components to diverge with a factor of $r^{-l}$, and the resulting dispersion equation is,

$$(ik_r)^2 + k_\theta^2 + k_\varphi^2 - i\left(\frac{\partial}{\partial r} + \frac{2}{r}\right)(ik_r) - i\frac{1}{r}\left(\frac{\partial}{\partial \theta} + \cot\theta\right)k_\theta - i\frac{1}{r\sin\theta}\frac{\partial}{\partial \varphi}k_\varphi = 0,$$

where $ik_r$, $k_\theta$, and $k_\varphi$ are wavevectors along $r$-, $\theta$-, and $\varphi$-directions, respectively, $k_r$ is a real number.

Among the three wavevector components, the $r$-component is purely imaginary, analogous to the imaginary transverse wavevector in plasmonics but free from metallic losses. As $r \to 0$, this component diverges, reflecting a field that decays rapidly as $r$ increases in any direction. This divergence drives all wavevector components governed by the dispersion equation to approach infinity, thereby enabling the electric field to achieve extraordinary localization in real space across all spatial dimensions.

*Narwhal-shaped wavefunction*—The mode volume quantifies the effective spatial region in which a given optical mode is confined. It is determined by integrating the electric energy density over all space and normalizing by the maximum value. Minimizing this volume requires the electric field to decay quickly from its peak, thereby concentrating the mode within a smaller region and enhancing the peak energy density.

Most optical modes follow a Gaussian-like electric field intensity profile, $I(x) = I_0 \exp\left(-\frac{x^2}{2\sigma^2}\right)$, where $I_0$ is the maximum intensity (located at $x=0$), and $\sigma$ characterizes the width of the mode profile (black curve in Fig. 1(b)). The relative rate of change of the intensity is $\left|\frac{d(\ln I)}{dx}\right| = \frac{|x|}{\sigma^2}$ (black curve in Fig. 1(c)). This expression shows that the relative rate of change increases as the distance ($|x|$) from the point of maximum intensity grows. However, near the region of maximum intensity, the relative rate of change is smaller, becoming zero exactly at the peak intensity ($x=0$). By comparison, an exponentially decaying evanescent wave has a position-independent (constant) relative rate of change, likewise restricting its effectiveness for minimizing mode volume.

To minimize the mode volume, a new wavefunction must be developed that maintains a high relative rate of change both near and far from the intensity peak. While Gaussian profiles exhibit rapid decay away from the center, their relative rate of change near the maximum is too low. In contrast, power-law functions (e.g., $|\mathbf{E}|^2 \propto r^{-2l}$) have a relative change rate of $\frac{2l}{r}$, which increases as $r$ decreases, resulting in faster intensity variation closer to the peak (red curves in Fig. 1(b) and 1(c)). However, the power-law function cannot be used as an eigen-wavefunction alone, as it diverges when integrated over the entire spatial domain.

We thus propose a novel hybrid wavefunction that transitions from a power-law form around the peak to an exponential tail at larger distances (Fig. 1(a)). By ensuring rapid decay from the high-intensity core to the far field, this hybrid profile significantly reduces the overall mode volume. Because of its resemblance to a narwhal's head, we term it the narwhal-shaped wavefunction.

*Experimental characterization of narwhal-shaped wavefunction*—We have developed a three-dimensional singular dielectric cavity operating in the microwave band (~1.3 GHz), enabling precise, direct, three-dimensional measurements of its eigen-wavefunction. Figure 2(a) shows the singular dielectric cavity, which comprises a three-dimensional biconical antenna and a twisted lattice cavity. Both components are made of dielectric materials: the twisted lattice cavity is formed from aluminum oxide, while the biconical antenna is fabricated from zirconium oxide.

In the twisted lattice cavity, the twist angle is 3.89°, and the lattice constant for both sets of identical photonic graphene lattices is 100 mm (~$\lambda/2$). The biconical antenna is composed of two zirconium oxide cones, separated by an air gap of approximately 0.02 mm (~$9 \times 10^{-5}\ \lambda$), enabling extreme electromagnetic field localization at the singularity. Zirconium oxide was selected for its higher dielectric constant, which increases the power-law exponent $l$, thereby amplifying the rate of change of the electric field in regions governed by the power-law profile.

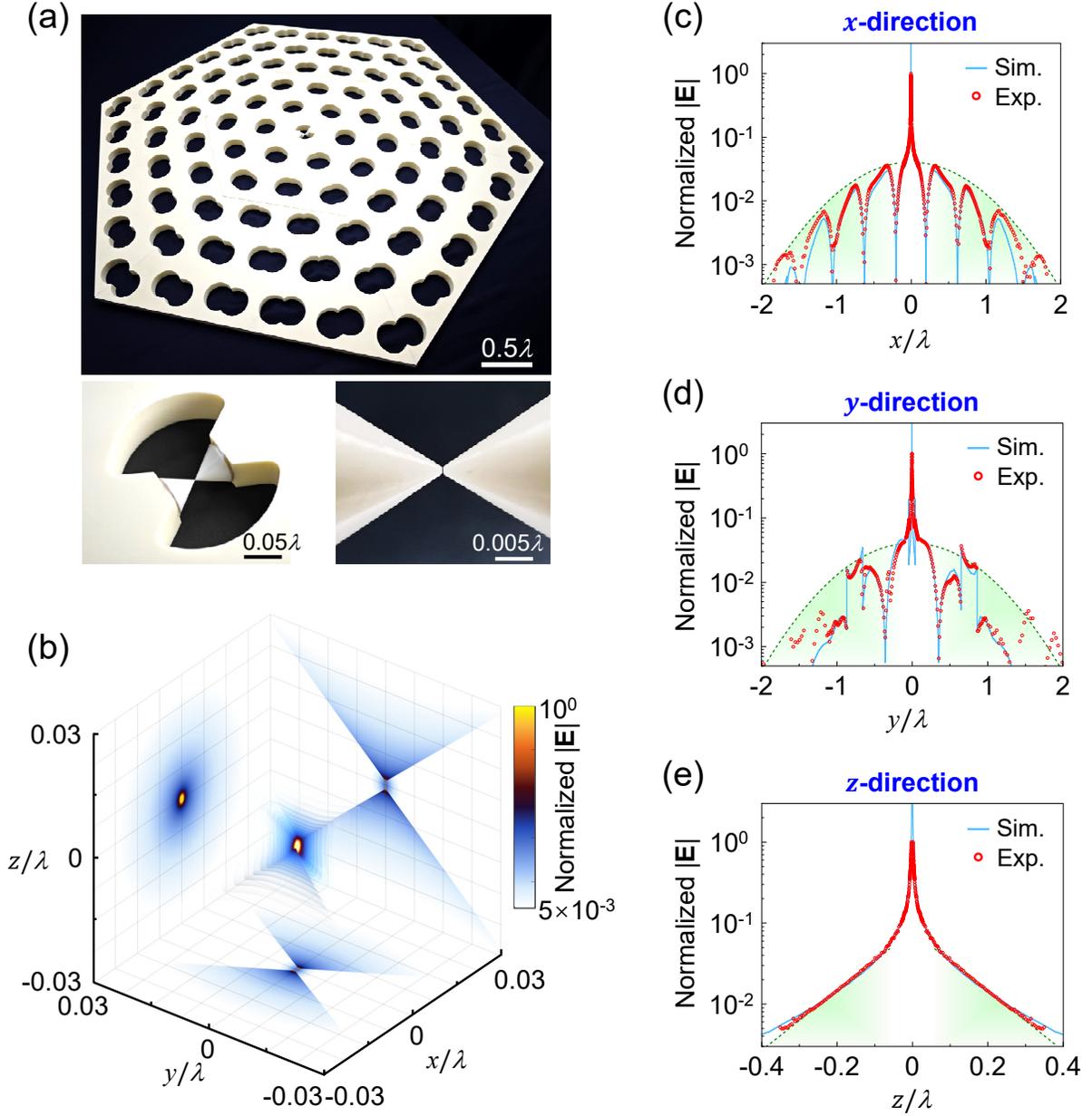

FIG. 2. Experimental characterization of a three-dimensional narwhal-shaped wavefunction. (a) Photograph of a singular dielectric cavity supporting a three-dimensional narwhal-shaped wavefunction. The cavity incorporates a biconical antenna and a twisted-lattice structure. Bottom insets: enlarged views of the biconical antenna region. (b) Three-dimensional full-wave simulation of the cavity's eigen-wavefunction, showing the mode's 3D intensity distribution along with its 2D projections on the $x$–$y$, $x$–$z$, and $y$–$z$ planes. (c)-(e) Cross-sectional intensity profiles taken through the singularity in each 2D projection. Solid lines: simulation results; dots: experimental measurements; dashed lines: Gaussian (c), (d) and exponential (e) functions for reference. The green-shaded regions indicate where the electric field decays exponentially.

Figures 2(b)-2(e) shows both the three-dimensional full-wave simulated and experimentally measured sub-diffraction-limited wavefunctions of the cavity. Figure 2(b) highlights the mode's three-dimensional intensity distribution, along with two-dimensional projections onto the $x$–$y$, $x$–$z$, and $y$–$z$ planes. Figures 2(c)-2(e) further presents cross-sectional intensity profiles (lines) taken through the singularity in each of the three projected intensity distributions, alongside the corresponding experimental results (dots). All three cross sections exhibit a distinct narwhal-shaped profile, and the experimentally measured intensity distribution matches well with the simulation.

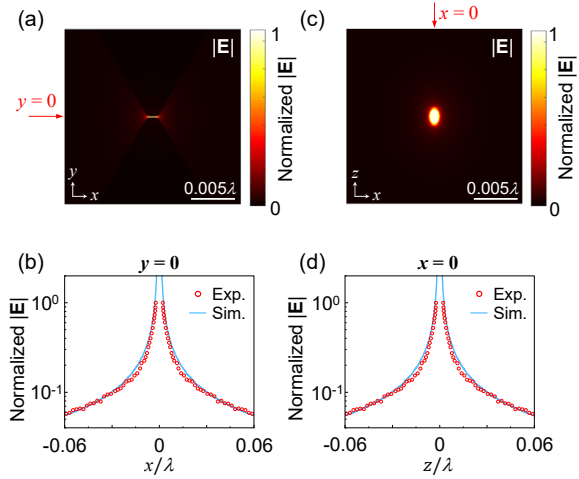

FIG. 3. Simulated and experimentally measured electric field intensity near the singularity. (a) Simulated electric field distribution of the narwhal-shaped wavefunction in the $x$–$y$ plane at $z=0$. (b) Comparison of simulated and experimentally measured electric field intensities at $y=0$ in (a). The electric field intensities are measured at $z=0.002\lambda$, as the probe cannot reach $z=0$. (c) Simulated electric field distribution in the $x$–$z$ plane at $y=0$. (d) Comparison of simulated and experimentally measured electric field intensities at $x=0$ in (c). Due to measurement limitations, the z-directed electric field is only recorded above the cavity ($z>0$), with values below ($z<0$) extrapolated from symmetry for reference.

Figure 3 compares the simulated and experimentally measured electric field intensity and phase distributions near the singularity. As shown in Figs. 3(b) and 3(d), the field intensity peaks at the singularity and decays outward following a power law of approximately $r^{-0.9}$, in excellent agreement with the simulations. This strongly localized field, which exhibits a power-law decay in all spatial directions, arises from the diverging imaginary radial wavevector.

In the two-dimensional cross section of this three-dimensional field, the phase varies sharply near $r=0$ along the $\varphi$-direction, indicating a diverging angular wavevector $k_\varphi$ (Fig. 4). Such rapid angular phase variation (diverging real angular wavevector) causes the pronounced radial field decay (imaginary radial wavevector), consistent with the dispersion equation. Figure 4(b) shows the directly measured phase shift around the singularity, again in excellent agreement with the simulations. Due to the unique narwhal-shaped wavefunction, the mode volume of the singular dielectric cavity is minimized. Three-dimensional full-wave simulations indicate that its mode volume is $5\times10^{-7}\ \lambda^3$, exceeding that of previously reported two-dimensional singular cavities by nearly three orders of magnitude [20, 23-25].

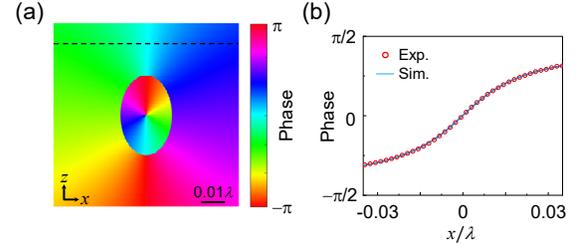

FIG. 4. Simulated and experimentally measured electric field phase distributions near the singularity. (a) Simulated phase distribution of the narwhal-shaped wavefunction in the $x$–$z$ plane at $y=0.02\lambda$. (b) Comparison of simulated and experimentally measured phases along the dashed line in (a).

*Conclusion*—In this work, we establish the foundational framework for sub-diffraction-limited dielectric photonics across all spatial dimensions. First, we introduce and experimentally validate a three-dimensional singular dispersion equation that enables the confinement of electromagnetic fields within dielectric singular cavities, operating below the diffraction limit in every spatial dimension. Second, we reveal that this exceptional confinement results from the unique narwhal-shaped wavefunction of dielectric singular fields, which exhibits a power-law decay near the point of maximum intensity and transitions to an exponential profile at larger distances. This hybrid wavefunction facilitates the rapid decay of the electric field from the peak intensity point throughout the spatial volume, effectively minimizing the mode volume. Third, we present the first direct experimental measurement of the wavefunction in a sub-diffraction-limited singular cavity, mapping the electric field distribution across the entire spatial domain. Our work enables the realization of photonic devices with exceptionally small mode volumes in lossless dielectric systems, offering significant promise for applications spanning physics, chemistry, biology, and engineering.

*Acknowledgements*—This work is supported by National Natural Science Foundation of China (grant nos. 12225402, 91950115, 11774014, 62321004, 92250302), national Key R&D Program of China (grant no. 2022YFA1404700), Beijing Natural Science Foundation (Z180011) and the New Cornerstone Science Foundation through the XPLORER PRIZE.

[1] P. A. M. Dirac, The quantum theory of the emission and absorption of radiation, Proc. R. Soc. Lond. A **114**, 243 (1927).
[2] E. M. Purcell, Spontaneous emission probabilities at radio frequencies, Phys. Rev. **69**, 681 (1946).
[3] K. J. Vahala, Optical microcavities, Nature (London) **424**, 839 (2003).
[4] S. Haroche, Nobel lecture: Controlling photons in a box and exploring the quantum to classical boundary, Rev. Mod. Phys. **85**, 1083 (2013).
[5] A. F. Kockum, A. Miranowicz, S. De Liberato, S. Savasta, F. Nori, Ultrastrong coupling between light and matter, Nat. Rev. Phys. **1**, 19 (2019).
[6] P. Forn-Díaz, L. Lamata, E. Rico, J. Kono, E. Solano, Ultrastrong coupling regimes of light–matter interaction. Rev. Mod. Phys. **91**, 025005 (2019).
[7] W. L. Barnes, A. Dereux, T. W. Ebbesen, Surface plasmon subwavelength optics, Nature (London) **424**, 824 (2003).
[8] E. Ozbay, Plasmonics: merging photonics and electronics at nanoscale dimensions, Science **311**, 189 (2006).
[9] J. N. Anker, W. P. Hall, O. Lyandres, N. C. Shah, J. Zhao, R. P. Van Duyne, Biosensing with plasmonic nanosensors, Nat. Mater. **7**, 442 (2008).
[10] J. A. Schuller, E. S. Barnard, W. Cai, Y. C. Jun, J. S. White, M. L. Brongersma, Plasmonics for extreme light concentration and manipulation, Nat. Mater. **9**, 193 (2010).
[11] H. A. Atwater, A. Polman, Plasmonics for improved photovoltaic devices, Nat. Mater. **9**, 205 (2010).
[12] D. K. Gramotnev, S. I. Bozhevolnyi, Plasmonics beyond the diffraction limit, Nat. Photon. **4**, 83 (2010).
[13] Y. Liu, X. Zhang, Metamaterials: a new frontier of science and technology, Chem. Soc. Rev. **40**, 2494 (2011).
[14] O. Hess, J. B. Pendry, S. A. Maier, R. F. Oulton, J. M. Hamm, K. L. Tsakmakidis, Active nanoplasmonic metamaterials, Nat. Mater. **11**, 573 (2012).
[15] M. Kauranen, A. V. Zayats, Nonlinear plasmonics, Nat. Photon. **6**, 737 (2012).
[16] M. S. Tame, K. R. McEnery, Ş. K. Özdemir, J. Lee, S. A. Maier, M. S. Kim, Quantum plasmonics, Nat. Phys. **9**, 329 (2013).
[17] J. B. Khurgin, How to deal with the loss in plasmonics and metamaterials, Nat. Nanotechnol. **10**, 2 (2015).
[18] K. L. Tsakmakidis, O. Hess, R. W. Boyd, X. Zhang, Ultraslow waves on the nanoscale, Science **358**, eaan5196 (2017).
[19] R. M. Ma, R. F. Oulton, Applications of nanolasers, Nat. Nanotechnol. **14**, 12 (2019).
[20] Y. H. Ouyang, H. Y. Luan, Z. W. Zhao, W. Z. Mao, R. M. Ma, Singular dielectric nanolaser with atomic-scale field localization, Nature (London) **632**, 287 (2024).
[21] S. Hu, S. M. Weiss, Design of photonic crystal cavities for extreme light concentration, ACS Photon. **3**, 1647 (2016).
[22] H. Choi, M. Heuck, D. Englund, Self-similar nanocavity design with ultrasmall mode volume for single-photon nonlinearities, Phys. Rev. Lett. **118**, 223605 (2017).
[23] S. Hu, M. Khater, R. Salas-Montiel, E. Kratschmer, S. Engelmann, W. M. J. Green, S. M. Weiss, Experimental realization of deep-subwavelength confinement in dielectric optical resonators, Sci. Adv. **4**, eaat2355 (2018).
[24] M. Albrechtsen, B Vosoughi Lahijani, R. E. Christiansen, V. T. H. Nguyen, L. N. Casses, S. E. Hansen, N. Stenger, O. Sigmund, H. Jansen, J. Mørk, S. Stobbe, Nanometer-scale photon confinement in topology-optimized dielectric cavities, Nat. Commun. **13**, 6281 (2022).
[25] A. N. Babar, T. A. S. Weis, K. Tsoukalas, S. Kadkhodazadeh, G. Arregui, B. Vosoughi Lahijani, S. Stobbe, Self-assembled photonic cavities with atomic-scale confinement, Nature (London) **624**, 57 (2023).
[26] R. M. Ma, Nanolaser technology with atomic-scale field localization, Nat Rev Electr Eng **1**, 632 (2024).
[27] J. B. Pendry, L. Martin-Moreno, F. J. Garcia-Vidal, Mimicking surface plasmons with structured surfaces, Science **305**, 847 (2004).
[28] N. Fang, H. Lee, C. Sun, X. Zhang, Sub-diffraction-limited optical imaging with a silver superlens, Science **308**, 534 (2005).
[29] K. L. Tsakmakidis, A. D. Boardman, O. Hess, 'Trapped rainbow' storage of light in metamaterials, Nature (London) **450**, 397 (2007).
[30] R. F. Oulton, V. J. Sorger, T. Zentgraf, R. M. Ma, C. Gladden, L. Dai, G. Bartal, X. Zhang, Plasmon lasers at deep subwavelength scale, Nature (London) **461**, 629 (2009).
[31] A. E. Miroshnichenko, S. Flach, Y. S. Kivshar, Fano resonances in nanoscale structures, Rev. Mod. Phys. **82**, 2257 (2010).
[32] A. I. Kuznetsov, A. E. Miroshnichenko, M. L. Brongersma, Y. S. Kivshar, B. Luk'yanchuk, Optically resonant dielectric nanostructures, Science **354**, aag2472 (2016).

[33] L. Feng, R. El-Ganainy, L. Ge, Non-Hermitian photonics based on parity–time symmetry, Nat. Photon. **11**, 752 (2017).
[34] T. Ozawa, H. M. Price, A. Amo, N. Goldman, M. Hafezi, L. Lu, M. C. Rechtsman, D. Schuster, J. Simon, O. Zilberberg, I. Carusotto, Topological photonics, Rev. Mod. Phys. **91**, 015006 (2019).
[35] H. Z. Chen, T. Liu, H. Y. Luan, R. J. Liu, X. Y. Wang, X. F. Zhu, Y. B. Li, Z. M. Gu, S. J. Liang, H. Gao, L. Lu, L. Ge, S. Zhang, J. Zhu, R. M. Ma, Revealing the missing dimension at an exceptional point, Nat. Phys. **16**, 571 (2020).
[36] X. R. Mao, Z. K. Shao, H. Y. Luan, S. L. Wang, R. M. Ma, Magic-angle lasers in nanostructured moiré superlattice, Nat. Nanotechnol. **16**, 1099 (2021).
[37] C. Wang, Z. Fu, W. Mao, J. Qie, A. D. Stone, L. Yang, Non-Hermitian optics and photonics: from classical to quantum, Adv. Opt. Photon. **15**, 442 (2023).
[38] H. Y. Luan, Y. H. Ouyang, Z. W. Zhao, W. Z. Mao, R. M. Ma, Reconfigurable moiré nanolaser arrays with phase synchronization, Nature (London) **624**, 282 (2023).
[39] R. M. Ma, H. Y. Luan, Z. W. Zhao, W. Z. Mao, S. L. Wang, Y. H. Ouyang, Z. K. Shao, Twisted lattice nanocavity with theoretical quality factor exceeding 200 billion, Fundam. Res. **3**, 537 (2023).
[40] S. T. Ha, Q. Li, J. K. W. Yang, H. V. Demir, M. L. Brongersma, A. I. Kuznetsov, Optoelectronic metadevices, Science **386**, eadm7442 (2024).
[41] J. M. Raimond, M. Brune, S. Haroche, Manipulating quantum entanglement with atoms and photons in a cavity, Rev. Mod. Phys. **73**, 565 (2001).
[42] S. Gleyzes, S. Kuhr, C. Guerlin, J. Bernu, S. Deléglise, U. B. Hoff, M. Brune, J.-M. Raimond, S. Haroche, Quantum jumps of light recording the birth and death of a photon in a cavity, Nature (London) **446**, 297 (2007).
[43] C.-Y. Lu, J.-W. Pan, Quantum-dot single-photon sources for the quantum internet, Nat. Nanotechnol. **16**, 1294 (2021).
[44] J. B. Khurgin, G. Sun, Comparative analysis of spasers, vertical-cavity surface-emitting lasers and surface-plasmon-emitting diodes, Nat. Photon. **8**, 468 (2014).
[45] J. B. Khurgin, Ultimate limit of field confinement by surface plasmon polaritons, Faraday Discuss. **178**, 109 (2015).